\documentclass[a4paper,11pt]{article}
\usepackage{pos}
\newcommand{\apj}{ApJ}
\newcommand{\aap}{AAP}
\newcommand{\araa}{ARAA}
\newcommand{\apjl}{ApJL}
\newcommand{\mnras}{MNRAS}

\newcommand{\prl}{Phys. Rev. Letters}

\title{Mass transfer stability for AM CVn binaries with white dwarf donors

}
\ShortTitle{Hot helium white dwarf mass transfer stability}

\author*[a,b,c]{Lucy O. McNeill}
\author[d,e,f]{Ryosuke Hirai}

\affiliation[a]{Department of Astronomy, Kyoto University, Kyoto 606-8502, Japan}
\affiliation[b]{Hakubi Centre for Advanced Research, Kyoto University, Kyoto 606-8317, Japan}
\affiliation[c]{Center for Interdisciplinary Theoretical and Mathematical Sciences (iTHEMS), RIKEN, Saitama 351-0198, Japan}

\affiliation[d]{Astrophysical Big Bang Laboratory (ABBL), Pioneering Research Institute (PRI), RIKEN, Saitama 351-0198, Japan}
\affiliation[e]{School of Physics and Astronomy, Monash University, Clayton, Victoria 3800, Australia}
\affiliation[f]{OzGrav: The ARC Centre of Excellence for Gravitational Wave Discovery, Australia}

\emailAdd{{mcneill@kusastro.kyoto-u.ac.jp}}
\emailAdd{ryosuke.hirai@monash.edu}

\abstract{
\textbf{Abstract}
Double white dwarf binaries (DWDBs) with Helium components are progenitors to AM CVn binary systems. Their Galactic production rate may be given by the {number} of stably mass transferring DWDBs in the Milky Way. {The theoretical criteria for stable mass transfer in DWDBs is calculated assuming that component white dwarfs are completely cold and degenerate. Respective fractions of surviving AM CVn and DWDB which merge are then calculated by applying this criteria to population synthesis estimates for Galactic DWDB.} 

However, emerging observations of the local DWDB population suggest that {Helium white dwarf (He WD) components} are typically hot, and only partially degenerate when they {begin} mass transferring. Using recent numerical simulations of {He WD} donors in DWDBs, we qualitatively describe a temperature dependent {stable mass transfer} criteria for Galactic DWDBs. {Mass transfer is even more stable than previously thought, or equivalently, DWDB mergers are even rarer. R}ealistic finite temperature treatments {will} deepen the dearth in observed AM CVn binaries compared to DWDB merger products. 

}

\FullConference{87th Fujihara Seminar: The 50th Anniversary Workshop of the Disk Instability Model in Compact Binary
Stars (DIM50TH2025), 22-26 September 2025, Tomakomai, Japan
}

\begin{document}
\maketitle
\section{Introduction}
Double white dwarf binaries (DWDBs) with Helium donors are the progenitors to a variety of transient phenomena and stellar populations {\cite{Webbink1984}}. DWDBs which are unstable to mass transfer and merge may be the origin of type .Ia supernova, calcium rich transients, R Cor Bor stars, and hot sub dwarf stars. On the other hand, DWDBs which stably mass transfer are potential progenitors of AM CVn binaries. 
\\

The observed rates of these transients and stellar populations can be compared with binary stellar evolution theory. For example, the AM CVn birth rate may be a proxy for the stable mass transfer rate in the Galactic DWDB population, or equivalently the survival rate of interacting DWDBs. This rate can ultimately be compared with the predicted population of white dwarfs in the Milky Way \citep{NelemansMT}. The instantaneous stellar radius response to mass loss is often used to characterize whether a given DWDB is dynamically stable or unstable to mass transfer. Assuming that white dwarfs (WDs) are completely degenerate at the moment of mass transfer \citep{Tutukov1996}, their radius is independent of internal thermal structure. In units of $R_\odot/100$ for white dwarfs with mass $m$, this zero--temperature mass radius relation is \citep{Verbunt1988}
\begin{equation}
\begin{split}
    & R_\mathrm{cold}(m) = 1.14 \left( \left(m/1.44 \right)^{-2/3}- \left(m/1.44 \right)^{2/3}\right)^{1/2} \times \\
    & \left(1+ 3.5 \left( m/ \left( 5.7 \times 10^{-4}\right)\right)\right)^{-2/3} + \left( \left( 5.7 \times 10^{-4}\right)/m \right)^{-2/3}.
    \label{eq:egg}
    \end{split}
\end{equation}
The mass transfer for a WD with $R_\mathrm{cold}$ will be stable if the rate of expansion of the Roche lobe radius {$R_\mathrm{RL}$}, given by \citep{Eggleton1983}
  \begin{equation}
R_\mathrm{RL} = a_\mathrm{RL}\frac{0.49 q^{2/3}}{0.6q^{2/3}+\mathrm{log} \left(1+q^{1/3} \right)},
\label{eq:RRL}
 \end{equation} 
 is faster than the expansion of the $R_\mathrm{cold}$. {Here, $a_\mathrm{RL}$ is the separation of a binary with mass ratio of the donor and accretor $q=m_D/m_A$.} Respective expansion rates of {$R_\mathrm{RL}$ and $R_\mathrm{cold}$} are quantified with differentials $\zeta$, where the response of $R_\mathrm{cold}$ to mass loss for a donor with mass $m$ and radius given by Equation~(\ref{eq:egg}) is characterised by:
\begin{equation}
\zeta_\mathrm{cold}\equiv\frac{\mathrm{d \ ln}R_\mathrm{cold}}{\mathrm{d \ ln}m} \approx -1/3.
\label{eq:zeta_cold}
\end{equation}

For the expansion rate of the binary Roche radius, the response to mass loss only depends on mass ratio $q$ \citep{Eggleton1983}
\begin{equation}
   \zeta_{\mathrm{RL}} \equiv \frac{\mathrm{d \ ln} \left( R_\mathrm{RL}/a_\mathrm{RL} \right)}{\mathrm{d \ ln}m} =  \frac{(q+1)}{3} \frac{2 \log
   \left(q^{1/3}+1\right)-{q^{1/3}}/({q^{1/3}+1})}{\left(0.6
   q^{2/3}+\log \left(q^{1/3}+1\right)\right)}.
\end{equation}
This can be approximated by $ \zeta_{\mathrm{RL}} \approx 1/3$ \citep{Pacynski1971}. 
There is a critical mass ratio, $q_\mathrm{crit}$ below which $| \zeta_{\mathrm{cold}} | < | \zeta_{\mathrm{RL}} |$. In this case, mass transfer is stable. This is given by \citep{Soberman1997} 
\begin{equation}
    q_\mathrm{crit} = 1 + \frac{\zeta_\mathrm{cold}-\zeta_\mathrm{RL}}{2} \approx 2/3.
    \label{eq:qcritcold}
\end{equation}
According to this dynamical mass transfer stability criteria for completely degenerate white dwarfs (Equation~(\ref{eq:qcritcold})), more equal mass ratios than $q>2/3$ lead to merger. Less equal mass binaries ($q<2/3$) are stable and can therefore survive as AM CVn binaries. 
\\

Analytic models have extended the stability boundary of Equation~(\ref{eq:qcritcold}) to include e.g. mass accretion destabilizing mass transfer. For example, \citep{NelemansMT} accounts for the possibility of super Eddington accretion, and predicts $q_\mathrm{crit}<2/3$, i.e. more mergers. Next, \citep{Marsh2004} account for angular momentum exchange facilitated by tidal torques, and find cases where it can stabilize, and also destabilize an otherwise stable, or unstable binary. The precise $q_\mathrm{crit}$ depends on the strength of tidal coupling and also the mode of mass transfer (disc or direct impact). But in any case, all scenarios result in $q_\mathrm{crit}\leq 2/3$. {Finally, these criteria can be applied to theoretical predictions for the Galactic DWDB population. For example, DWDB component masses from Galactic population synthesis are shown in \citep{NelemansMT}'s Fig. 1 and \citep{Kremer2017}'s Fig 7. The most populated region in $m_D$, $m_A$ space is just below the theoretical $m_D/m_A=q_\mathrm{crit}=2/3$ boundary, with $m_D$ (donor) roughly between 0.18-0.26~$M_\odot$ and $m_A$ (accretor) between 0.30-0.50$M_\odot$. In other words, most predicted Galactic DWDBs are low $q$ systems with $q<2/3$ which survive as AM CVn systems.} 
\\

This is not consistent with observations. For example, the rates of AM CVn given by the ELM (Extremely Low Mass) WD survey \citep{Kilic2016} are a factor of 10 smaller than the rate of DWDB mergers \citep{Brown2016}. Assuming that we understand selection effects associated with calculating these rates, it is challenging to mediate this discrepancy if the interacting DWDB survival rate is given by Equation~(\ref{eq:qcritcold}), and if population synthesis estimates for DWDB are accurate. One explanation for relatively low AM CVn birth rates could be that most DWDBs can not survive as AM CVn due to the accretor (not donor) response \citep{Shen2015}. 
\\

Recent advances in the detection of detached DWDBs by eclipsing methods \citep{Burdge2020} reveal a hot and large population of He WD in short period DWDB which will interact within a Hubble time (orbital period $P_\mathrm{orb} \lesssim$ 1 hr). If the Galactic population of DWDBs typically has hot and large donors at the onset of mass transfer \citep{MH25}, then the basic assumption leading to Equation~(\ref{eq:qcritcold})-- that Helium white dwarfs have cooled to be degenerate (Equation~(\ref{eq:egg})) at the onset of mass transfer may not be valid.
\\

There are also recent exciting discoveries of DWDBs undergoing mass transfer. In particular, ZTF J0127+5258 \citep{Burdge2023} is the first ever detection of a mass transferring eclipsing DWDB with the donor temperature measured, which is 16,400~K. In addition, its measured period derivative $\dot{P}_\mathrm{orb}$ is negative. In principle, the measured $\dot{P}_\mathrm{orb}$ can be used in conjunction with other observables (e.g. radial velocity curves) to constrain binary properties and the mass transfer rate. In Table~\ref{tab:WD1} we show two cases for the inferred properties of ZTF J0127+5258. The first row corresponds to the case of a $ 0.15 + 0.75 M_\odot$ binary used for the light curve fitting. The second row corresponds to the inferred masses assuming a mass transfer rate, given the measured period derivative. 
\begin{table*}
\caption{Properties of ZTF J0127+5258 from \cite{Burdge2023}. The first row are the masses and {donor} radius assumed for light curve fitting, assuming that the donor fills its Roche Lobe and that the accretor has a disc. The second row is the inferred masses and donor radius, combining binary evolution models for mass loss and the measured orbital period derivative. 
}
\label{tab:WD1}
\begin{center}
\begin{tabular}{ c c c c c }
\hline
\textbf{Mass estimation case} & {$m_D$} ($M_\odot$) & {$m_A$} ($M_\odot$) & $R$ ($R_\odot$)& $T_\mathrm{eff}$(K)  \\
 \hline
    Light curve fit (assumption) &  0.15 &  $0.75 $ ($q=0.2$) & 0.046 & 16,400   \\
        High mass transfer rate (inference) & $0.19 \pm{0.03}$  &  $0.75\pm{0.11}$ ($q=0.25$)  & 0.051 & 16,400  \\
\hline
\end{tabular}
\end{center}
\end{table*}
Only the second row utilizes the $\dot{P}_\mathrm{orb}$ measurement.
\\

Given these recent observational and theoretical developments, we seek a finite temperature theoretical framework to understand mass transfer stability of DWDBs with Helium white dwarf donors. In Section \ref{sec:finite-wd} we compare the inferred donor mass measurements of ZTF J0127+5258 with theoretical models. We show that mass transfer is more stable for finite temperature white dwarfs. In Section \ref{sec:discussion} we discuss implications for AM CVn birth rates summarise future work.
\section{Finite temperature Helium white dwarfs}
\label{sec:finite-wd}
\begin{figure}
  \centering
    \includegraphics[width=0.8\textwidth]{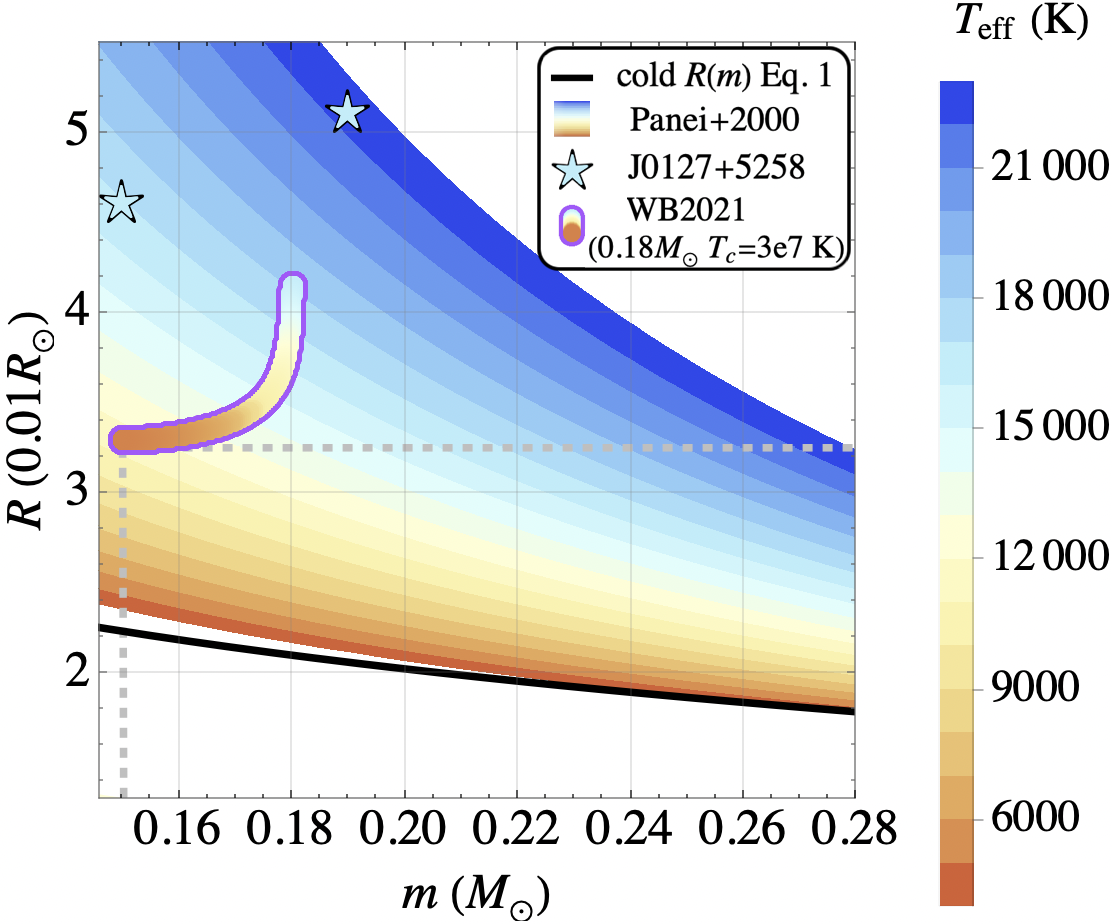}
  \caption{Temperature in K as a function of mass in $M_\odot$ (horizontal axis) and radius in $0.01 R_\odot$ (vertical axis) for a variety of theoretical stellar models and observations of He WD donors. The contours in the background approximate the finite temperature single He WD models from \cite{Panei2000}. The thick track with the purple outline is the $0.18M_\odot$ donor from \cite{Wong2021} being stripped to $0.15M_\odot$ by stable mass transfer, which is referred to as Md18-Tc3e7-Ma0p75. Notably, unlike a completely degenerate WD, the donor shrinks as a response to mass transfer. The two stars are different candidate properties of J0127+5258 in \cite{Burdge2023}. Compared to the single He WD models, both the mass transfer model (track) and candidate J0127+5258 donors (stars) are cooler for a given mass and radius.
  }
\label{fig:Sc-plot}
\end{figure}

The key ingredient for mass transfer stability in Equation~\ref{eq:qcritcold} is the donor's radius response to mass transfer, which is conventionally characterised by the cold mass--radius relation Equation~(\ref{eq:egg}) \citep{Soberman1997,NelemansMT}. When white dwarfs have hot surface temperatures $\gtrsim 10,000$~K, {their surface layers are no longer fully degenerate and are partially supported by thermal pressure. Therefore, the stellar radius is no longer a simple function of mass, but will also depend on the thermal structure and mass transfer timescale, which in turn depends on details of the energy transport and mass transfer mechanisms.} 
Then, the response of the finite temperature WD to mass loss
\begin{equation}
\zeta\equiv\frac{\mathrm{d \ ln}R}{\mathrm{d \ ln}m}
\label{eq:zeta}
\end{equation}
must be solved according to how the donor's thermal properties change, rather than Equation~(\ref{eq:zeta_cold}) which assumes that this response is independent of thermal properties via Equation~(\ref{eq:egg}). For detailed white dwarf energy transport models, we refer to  \cite{Panei2000}, who solve the equilibrium structure of pure helium WD models (without the hydrogen layer) for a variety of surface temperatures. These are the dashed lines in \cite{Panei2000}'s finite temperature mass--radius relations in their Figure 3. These hydrogen-free models may be representative of WD at the onset of mass transfer, in the sense that they have been stripped of their diffuse envelope. The best fit power law approximation for these models is given by:
\begin{equation}
R(m,T_\mathrm{eff}) = 0.0104 \times 10^{-0.0622 T_\mathrm{eff}^{1.5}} \times m^{-0.384 - 0.258 T_\mathrm{eff}^{1.5}} R_\odot,
\label{eq:RMTPanei2}
\end{equation}
with $T_\mathrm{eff}$ is in units of 10,000~K, and $m$ is in units of $M_\odot$. We plot these in the blue-red contours in Figure~\ref{fig:Sc-plot}. 
The plotting window from \cite{Panei2000} is outlined by the grey dashed lines, where we have extrapolated into the upper half.
We note that these are single stellar models in hydrostatic and thermal equilibrium. Therefore they may not describe the temperature of a given mass $m$ and radius $R$ of a He WD during mass transfer, since the He WD {is not be able to maintain thermal equilibrium, especially in the nondegenerate outer regions}.
It will be instructive to compare these models with observations and theoretical models of He WDs undergoing mass transfer. 
\\

More recently, stellar evolution models of finite temperature He WD also include the effects of mass transfer from a He WD donor \cite{Wong2021}. Unlike \citep{Panei2000}, these stellar models are generally not in {thermal equilibrium due to the relatively short mass transfer timescale.} 
We take the initially $m=0.18M_\odot$, $T_c=3 \times 10^7$~K donor with a point mass $0.75M_\odot$ companion, which we refer to as Md18-Tc3e7-Ma0p75. We plot its trajectory in mass--radius space with a purple outline (starting from upper right and moving down to the bottom left) in Figure~\ref{fig:Sc-plot}. The corresponding surface temperature is shown by the red-blue colour bar. As the initially $T_\mathrm{eff}=12,800$~K, $m=0.18 M_\odot$ donor is reduced to $0.15 M_\odot$, the central entropy $S_c$ remains almost constant (within 0.1$\%$). 
\\

The most notable feature of the Md18-Tc3e7-Ma0p75 model is that the donor {rapidly} shrinks during mass transfer. This is in contrast to a completely degenerate star, which would expand during mass transfer according to Equation~(\ref{eq:zeta_cold}).
{In addition, the donor surface temperature is always cooler (= the surface luminosity is lower) than the equivalent equilibrium WD model of a given mass and radius computed by \cite{Panei2000}. The donor is losing mass on a timescale much shorter than the thermal timescale, and in fact we find that the central entropy remains roughly constant throughout the mass transfer evolution. This rapid mass loss causes a drop in the surface luminosity and thus the deviation in surface temperature from the \cite{Panei2000} models become more pronounced towards later times (lower-mass end of the track).}
\\

To make a comparison with observations, in Figure~\ref{fig:Sc-plot} we also plot the two inferred mass and radii for ZTF J0127+5258 (Table~\ref{tab:WD1}) with the star markers. Both cases have $T_\mathrm{eff}$ cooler than the underlying single WD model contours. This is a similar trend to the model of \cite{Wong2021}. Between the two, the $0.15M_\odot$, $0.046 R_\odot$ candidate's surface temperature is closer to the corresponding single WD model. 

\section{Discussion and summary}
\label{sec:discussion}
We have discussed finite temperature He WD donors in mass transferring double white dwarf binaries, in both the observational and theoretical context. In the case of hot ($T_\mathrm{eff}>10,000$~K) low mass ($m \lesssim 0.2 M_\odot$) donors, He WDs shrink as a response to mass transfer according to Equation~(\ref{eq:zeta_finite}). It is unclear whether massive and/or cooler donors also shrink in response to mass transfer. But at the very least, finite temperature white dwarfs will expand less than the cold relation (Equation~(\ref{eq:zeta_cold})). The radius response of finite temperature white dwarfs is therefore
\begin{equation}
\zeta_S = \left. \frac{\mathrm{d \ ln  }  R}{\mathrm{d \ ln } m} \right|_{\substack{S_\mathrm{c}}} > -1/3 
\label{eq:zeta_finite}
\end{equation}
Replacing $\zeta_\mathrm{cold}$ with $\zeta_S$ in Equation~(\ref{eq:qcritcold}), if $\zeta_S>-1/3$, then the critical mass ratio at the stability boundary is given by
\begin{equation}
q_\mathrm{crit} > 2/3.
\end{equation}
In other words, only very equal mass DWDBs are unstable to mass transfer according to this simple criteria. In this case, the observed Galactic AM CVn birth rate is even smaller than the theoretical expectations by \cite{NelemansMT,Kremer2017}. These results only widen the discrepancy between the ELM survey results of \cite{Kilic2016,Brown2016} and binary stellar evolution theory for DWDBs.
\\

Future space-based gravitational wave detectors, such as the Laser-Interferometer-Space-Antenna (LISA) \citep{LISA} will be capable of detecting 10-100's of thousands of Galactic double white dwarf binaries with orbital periods <20 minutes, including mass transferring AM CVn systems. These are unprecedented numbers compared to optical surveys e.g. the ELM survey. It will therefore be possible to precisely measure the rates of stable compared to unstable mass transfer in the short period DWDB population \citep{Seto2022}. Until then, it is of paramount importance to improve the theoretical understanding of binary evolution in Galactic DWDB to explain the apparent dearth in Galactic AM CVn, and create accurate models for $\dot{P}_\mathrm{orb}$ due to mass transfer {in individual interacting DWDBs}.

\end{document}